\documentclass[a4paper]{article}

\usepackage{INTERSPEECH2022}

\usepackage[dvipsnames]{xcolor}
\usepackage{tipa}
\usepackage{textgreek}
\usepackage{hyperref}
\usepackage{graphicx}
\usepackage{tikz}

\usepackage[symbol]{footmisc}

\title{Deep Speech Synthesis from Articulatory Representations}
\name{Peter Wu$^1$, Shinji Watanabe$^2$, Louis Goldstein$^3$, Alan W Black*$^2$, Gopala K. Anumanchipalli*$^1$\thanks{*Equal advising.}}
\address{
  $^1$University of California, Berkeley, United States\\
  $^2$Carnegie Mellon University, United States\\
  $^3$University of Southern California, United States}
\email{peterw1@berkeley.edu}

\begin{document}

\maketitle
\begin{abstract}
  In the articulatory synthesis task, speech is synthesized from input features containing information about the physical behavior of the human vocal tract. This task provides a promising direction for speech synthesis research, as the articulatory space is compact, smooth, and interpretable. Current works have highlighted the potential for deep learning models to perform  articulatory synthesis. However, it remains unclear whether these models can achieve the efficiency and fidelity of the human speech production system. To help bridge this gap, we propose a time-domain articulatory synthesis methodology and demonstrate its efficacy with both electromagnetic articulography (EMA) and synthetic articulatory feature inputs. Our model is computationally efficient and achieves a transcription word error rate (WER) of 18.5\% for the EMA-to-speech task, yielding an improvement of 11.6\% compared to prior work. Through interpolation experiments, we also highlight the generalizability and interpretability of our approach.
\end{abstract}
\noindent\textbf{Index Terms}: speech synthesis, articulatory synthesis

\section{Introduction}
\label{sec:intro}

Speech synthesis has seen rapid development in recent years with deep learning based techniques. These models have shown success in text-to-speech (TTS) \cite{Wang2017Tacotron, hayashi2021espnet2, prenger2019waveglow}, speech-to-speech translation (S2ST) \cite{tjandra2019speechtranslation, jia2019speechtranslation, inaguma2020espnetspeechtranslation}, voice conversion (VC) \cite{polyak21facebookresynthesis, wu2021privacy, sisman2020overview}, and tasks with other modalites \cite{anumanchipalli2019speech, yu2019durianmultimodalsynthesis, gaddy-klein-2021-improved}. Moreover, this technology has yielded impactful technologies like speech synthesis aids for people with blindness or paralysis \cite{karmel2019speechsynthesisforblind, angrick2019ecogbraintospeech, anumanchipalli2019speech}. While speech synthesizers have already shown promising results in multiple domains, technologies like brain-to-speech devices remain challenging in build \cite{anumanchipalli2019speech}. Thus, these unsolved tasks require new algorithms in order to achieve the development of high-fidelity, open-vocabulary synthesizers. To this end, our work focuses on devising a deep speech synthesis methodology that is computationally efficient, real-time, and high-fidelity. We propose a time-domain articulatory synthesis approach that is suitable for attaining these three properties and empirically validate our method on two distinct articulatory modalities, EMA and a synthetic articulatory space. Our deep learning models also exhibit valuable interpretability properties, which we demonstrate through interpolation experiments. Audio samples, code, and additional related information are available at \href{https://articulatorysynthesis.github.io}{https://articulatorysynthesis.github.io}.

\section{Speech Synthesis}
\label{sec:relatedworksspeechsynthesis}

\subsection{Deep Speech Synthesis}
\label{sec:deepspeechsynthesis}

Currently, state-of-the-art speech synthesis algorithms use deep learning \cite{hayashi2021espnet2, anumanchipalli2019speech, jia2021translatotron, polyak21facebookresynthesis, gaddy-klein-2021-improved}. While existing methods can generate high-fidelity speech, they tend to be computationally expensive and difficult to interpret and generalize \cite{Nekvinda2020multilingualtts, zhang2019multilingualtts}. We attribute underspecification to the primary cause of these issues, as speech data is very high dimensional and current algorithms lack sufficient inductive biases. To help bridge this gap, we devise deep articulatory synthesis techniques that exhibit suitable computational efficiency, generalizability, and interpretability properties by behaving more similarly to the human speech production process than existing methods.

\subsection{Articulatory Synthesis}

Articulatory synthesis generally refers to the task of synthesizing speech from articulatory features, i.e., features containing information about the physical behavior of the human vocal tract \cite{fant1991articulatorysynthesis, rubin1981articulatorysynthesis, scully1990articulatorysynthesis, jiachen2022gesture}. We identify two primary research directions in articulatory synthesis: 1. modelling the human vocal tract \cite{fant1995lfvocaltractmodal, iskarous2003haskinscasy, birkholz2013vtl}, and 2. learning the mapping from articulatory features to speech through a statistical means \cite{aryal2016datadrivendnn, bocquelet2014dnn, chen2021ema2s}. The former direction, due to its focus on computational modelling, has yielded articulatory synthesizers that are interpretable and relatively space-efficient but computationally slow. On the other hand, the latter direction has yielded methods that are much faster but have worse interpretability and memory efficiency. Ideally, speech synthesizers should have low space and time complexities, which would enable many impactful real-time applications. For example, such systems could allow patients with paralysis or aphasia to communicate naturally at any moment in time. Thus, we focus on making methods in the second research direction more memory-efficient in this work. Additionally, we highlight how statistical articulatory synthesis methods could also be highly interpretable, thus containing all of the benefits of articulatory synthesizers built using physical modelling.

Another motivation for our statistical research direction is the transferability of our methodology to all forms of speech synthesis. Current state-of-the-art speech synthesis systems rely on an intermediate speech representation, typically a spectrum or a learned representation \cite{kong2020hifigan, morrison2022cargan, badlani2021onettsalignment, kim2021vits, Elias2021ParallelTacotron2}. Inductive biases offer one potential way of making these models efficient, generalizable, and interpretable, as mentioned in Section \ref{sec:deepspeechsynthesis}. Constraining these intermediate representations to an articulatory feature space is one way to impose such an inductive bias, since a limited set of articulator configurations can completely specify all possible human speech \cite{nam2012procedure}. The resulting model would then need to perform an articulatory-to-speech mapping, of which the behavior is relatively unknown to our knowledge. This work aims to bridge this gap by studying the efficiency, generalizability, interpretability, and fidelity of such a mapping using two distinct articulatory modalities. Specifically, we use the MNGU0 EMA dataset \cite{richmond2011mngu0} and another corpus generated by VocalTractLab \cite{birkholz2013vtl}, detailed in Section \ref{sec:datasets}.

While deep EMA-to-speech models have been previously studied, as far as we are aware \cite{taguchi2018articulatory, stone2020ema, liu2018articulatory}, current models are not highly intelligible, achieving a transcription WER of around 30\% on open-vocabulary tasks \cite{taguchi2018articulatory}. In this work, we build an EMA-to-speech model that achieves a transcription WER of 18.5\% and perform detailed error analyses on the synthesized utterances. We also extend this approach to building a speech synthesizer using a synthetic articulatory modality. This model is efficient, high-fidelity, and interpretable, which has previously been unattained to our knowledge. We detail these models and our proposed time-domain articulatory synthesis methodology in Section \ref{sec:models} below.

\section{Deep Articulatory Models}
\label{sec:models}

\subsection{Frequency- and Time-Domain Modeling}

Similarly to the state-of-the-art speech synthesis works discussed in Section \ref{sec:relatedworksspeechsynthesis}, current deep articulatory synthesis works rely on synthesizing an intermediate spectrum representation, from which waveforms are generated \cite{Csapo2020UltrasoundbasedAM, georges2020towards}. Since this behavior is not present in the human speech production process, we propose a model that directly maps articulatory representations to waveforms. We refer to this approach as a time-domain one, as it does not explicitly rely on a frequency-based intermediate. This modification noticeably improves model efficiency while achieving comparable intelligibility, as discussed in Sections \ref{sec:computational_efficiency} and \ref{sec:synthesis_quality}. We detail our spectrum-intermediate baseline in Section \ref{sec:spectrumintermediatebaseline} and our time-domain method in Section \ref{sec:time_domain_hifi}.

\subsection{Spectrum-Intermediate Baseline}
\label{sec:spectrumintermediatebaseline}

For our baseline, we build on a state-of-the-art model proposed by Gaddy and Klein \cite{gaddy-klein-2021-improved}. Namely, we map articulatory representations to spectrums using a six-layer Transformer \cite{Vaswani2017attentiontransformer} prepended with three residual convolution blocks. To map spectrums to waveforms, we use HiFi-GAN \cite{kong2020hifigan}, which we make autoregressive using the audio encoder from CAR-GAN \cite{morrison2022cargan}. For our spectrum representation, we use Mel spectrograms instead of MFCCs, as done in the CAR-GAN paper and most deep speech synthesis works \cite{morrison2022cargan, Wang2017Tacotron, hayashi2021espnet2}.
We omit the phonemic loss to avoid requiring phoneme annotations during training and instead improve model performance by adding the HiFi-GAN adversarial loss \cite{kong2020hifigan}. Since articulatory representations in this work are pre-aligned with waveforms, we also omit the dynamic time warping loss. We refer to this model as the spectrum-intermediate (Spec.-Int.). Further modeling details can be found in the accompanying codebase.

\subsection{HiFi-CAR Model}
\label{sec:time_domain_hifi}

For our time-domain model, we feed our articulatory input features directly into HiFi-GAN \cite{kong2020hifigan}, which we make autoregressive using the audio encoder from CAR-GAN \cite{morrison2022cargan}. To our knowledge, directly feeding articulatory inputs into a deep vocoder architecture has not yielded any successful results previously. However, we observe that this model is comparable to our baseline, as discussed in Section \ref{sec:synthesis_quality}. Moreover, removing the need for an articulatory-to-spectrum architecture noticeably improves computational efficiency, as discussed in Section \ref{sec:computational_efficiency}. We refer to this model as HiFi-CAR below. Further modeling details can be found in the accompanying codebase.

\section{Datasets}
\label{sec:datasets}

\subsection{Electromagnetic Articulography (EMA)}
\label{sec:ema_mngu0_dataset}

For our first task, we perform EMA-to-speech using the MNGU0 dataset \cite{richmond2011mngu0}. MNGU0 contains 67 minutes of 16 kHz, single-speaker speech collected with the speaker instrumented in an EMA machine providing 200 Hz samples of EMA features. These 12-dimensional features contain the midsagittal x and y coordinates of jaw, lip, and tongue positions. We use the original train-test split with 1,129 training utterances and 60 test ones, and randomly choose 60 training datapoints for the validation set. Since EMA features do not contain voicing information, we concatenate them with estimated F0 sequences extracted using CREPE \cite{kim2018crepe, morrison2022cargan}.

\begin{figure}[t]
  \includegraphics[width=77mm]{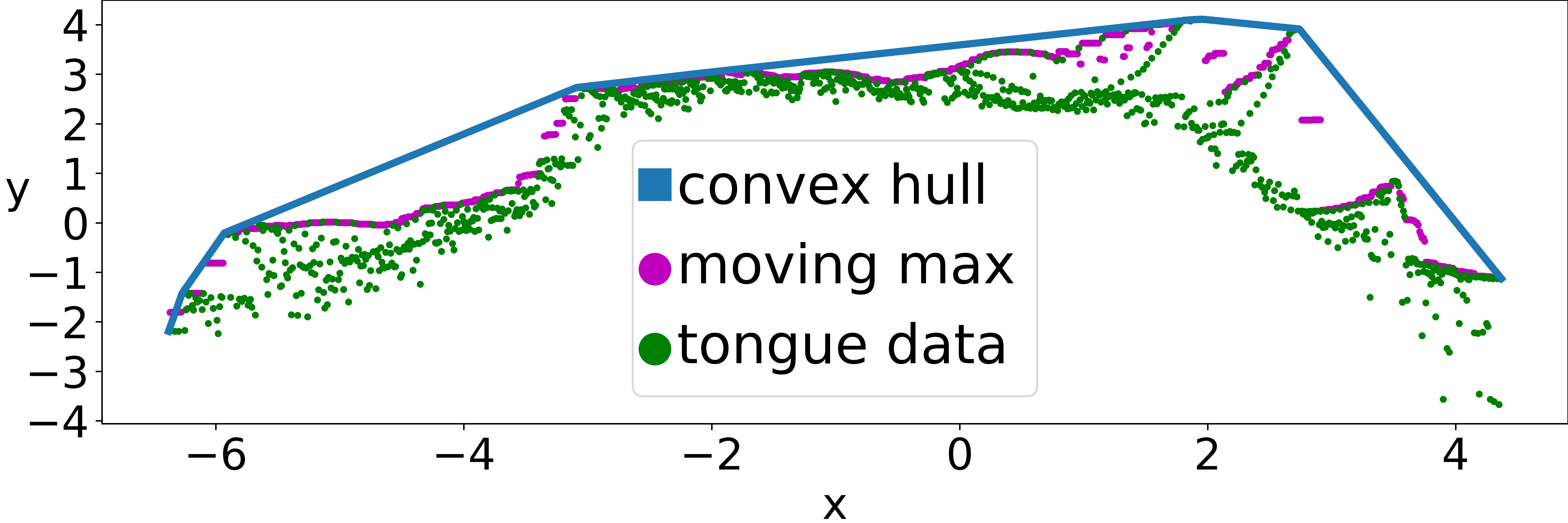}
  \caption{Estimating the palate location with a convex hull and a moving maximum on tongue data in the training set (Sec. \ref{sec:ema_mngu0_dataset}).}
  \label{fig:roof_distances}
\end{figure}

Palate information, which is important for determining consonant sounds
\cite{hagedorn11_interspeech}, is also not directly present in EMA. Thus, we estimate the location of the palate using tongue data in the training set using two methods. First, we compute the convex hull of the tongue coordinates \cite{nam2012procedure}. Since this estimate does not account for the concave portions of the tongue, we also compute the moving maximum along the x-axis, using a window size of 32. For each EMA feature sequence, we estimate the tongue-to-palate distance by subtracting the tongue tip, body, and dorsum y-coordinates from both palate estimates, yielding 6 additional features. Figure \ref{fig:roof_distances} plots our tongue training data and palate estimates on MNGU0's normalized xy-coordinates.

\subsection{Synthetic Articulatory Features}
\label{sec:birk_dataset}

Since EMA data does not explicitly contain enough manner information to perfectly reconstruct the original speech \cite{anumanchipalli2019speech}, we also experiment with synthetic articulatory data that does. Namely, we use the vocal tract model from Birkholz et al. \cite{birkholz2013vtl} to create a single-speaker corpus of pseudo-words, each composed of two to three vowel and consonant sounds. Our training set has 10,000 such utterances, and our validation set has 250, totaling a few hours of speech. For our test set, we use vocal tract model outputs corresponding to the first 99 phoneme sequences in the CMU US KAL Diphone database \cite{lenzo2003kal}. All waveforms have a sampling rate of 44100 Hz and 30-dimensional articulatory features are recorded every 110 samples. We refer to this dataset as the Birkholz-Pseudoword (Birk.-Pseudo.) dataset below.

\section{Computational Efficiency}
\label{sec:computational_efficiency}

Computational efficiency during inference is essential for building real-time, on-device speech synthesizers. We observe that our time-domain articulatory synthesis model is more time- and space-efficient than the spectrum-intermediate baseline. Table \ref{space_complexity} contains the EMA-to-speech inference times and number of parameters for both models. GPU trials use one RTX 2080 Ti GPU, and CPU trials use none. We report inference time as the mean and standard deviation of five trials, each calculating the average time to synthesize an utterance in our test test. Our time-domain model is almost $\bm{2\times}$ faster than the baseline in the CPU-only case and uses $\bm{7\times}$ less parameters. Our synthetic articulatory experiments, which use similar hyperparameters, yield matching trends, detailed in the supplementary material linked in Section \ref{sec:intro}. These results suggest that directly mapping articulatory features to speech is more efficient than relying on an intermediate spectral representation.

\begin{table}[t]
\centering
\begin{tabular}{lccc}
\hline
\textbf{Data} & \textbf{CPU (s)} \textcolor{ForestGreen}{$\downarrow$} & \textbf{GPU (s)} \textcolor{ForestGreen}{$\downarrow$} & \textbf{Params.} \textcolor{ForestGreen}{$\downarrow$}\\
\hline
HiFi-CAR & $\mathbf{4.10 \pm 0.03}$ & $\mathbf{0.37 \pm 0.00}$ & $\mathbf{1.3*10^7}$ \\
Spec.-Int. & $7.61 \pm 0.02$ & $0.39 \pm 0.00$ & {$9.5*10^7$} \\
\hline
\end{tabular}
\caption{\label{space_complexity}
Average inference time and number of parameters for EMA-to-speech models. See Section \ref{sec:computational_efficiency} for details.
}
\end{table}

\section{Interpolation}
\label{sec:interpolation}


\subsection{Vowel Interpolation}
\label{sec:vowel_interpolation}

\begin{figure}[t]
  \includegraphics[width=77mm]{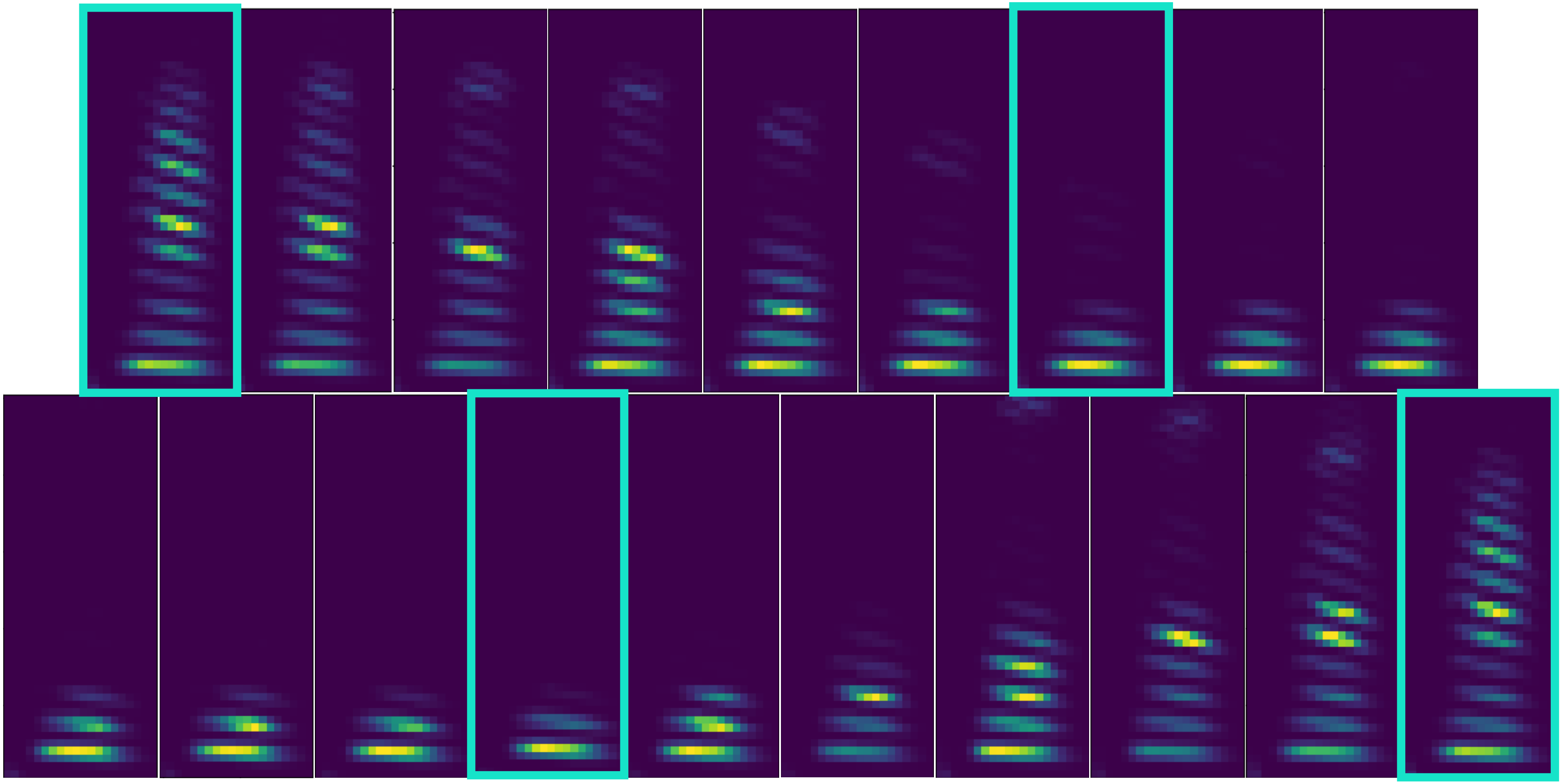}
  \caption{Vowel interpolation between "ta" (top left), "tu", "ti", and "ta" (bottom right), with these four diphones boxed.}
  \label{fig:vowel_grad}
\end{figure}

To study the generalizability of our time-domain model, we perform interpolation experiments, prompting our model to synthesize unseen articulatory representations between pairs of sounds. All generations are available through our accompanying \href{https://articulatorysynthesis.github.io}{link}. First, we interpolate between ``ta'', ``tu'', and ``ti'' to analyze how well our model generalizes across vowels. Similarly to our synthetic articulatory task, we generate the articulatory features for these sounds using VocalTractLab \cite{birkholz2013vtl}, which can provide versions of these utterances with the same duration. For each of the three possible pairs of sounds, we perform a linear interpolation between the two articulatory features, generating seven evenly spaced weighted combinations. Figure \ref{fig:vowel_grad} contains the mel-spectrograms of the generated speech from our model for each of these combined articulatory features. The transitions between spectrum values in each interpolation are smooth, suggesting that our network is able to model the continuity of articulator movements, at least with respect to vowels. This trend is also reflected in our listening test, where English-speaking, MTurk listeners classified each of the seven ``ta''$\rightarrow$``tu'' utterances as ``ta'', ``t\textepsilon'', ``tu'', or ``possibly'' followed by one of these three syllables. Each utterance received 10 votes, yielding 70 annotations by a total of 19 listeners. To calculate which syllable the listeners assigned to each utterance on average, we convert each vote to a size-3 one-hot vector, multiplied by 0.5 if ``possibly'' was used. Index 0 is non-zero if ``ta'' was chosen, 1 if ``t\textepsilon'', and 2 if ``tu''. For each utterance, we plot the mean vector as three adjacent bars, one for each scalar, depicted in Figure \ref{fig:interp_bar_plot}. In this transition from ``ta'' to ``tu'', listeners steadily perceive ``ta'' less and ``tu'' more, with ``t\textepsilon'' peaking in the middle, matching human vowel speech production behavior \cite{alfonso1982dynamics}.

\subsection{Consonant Interpolation}
\label{sec:consonant_interpolation}

\begin{figure}[t] 
  \includegraphics[width=77mm]{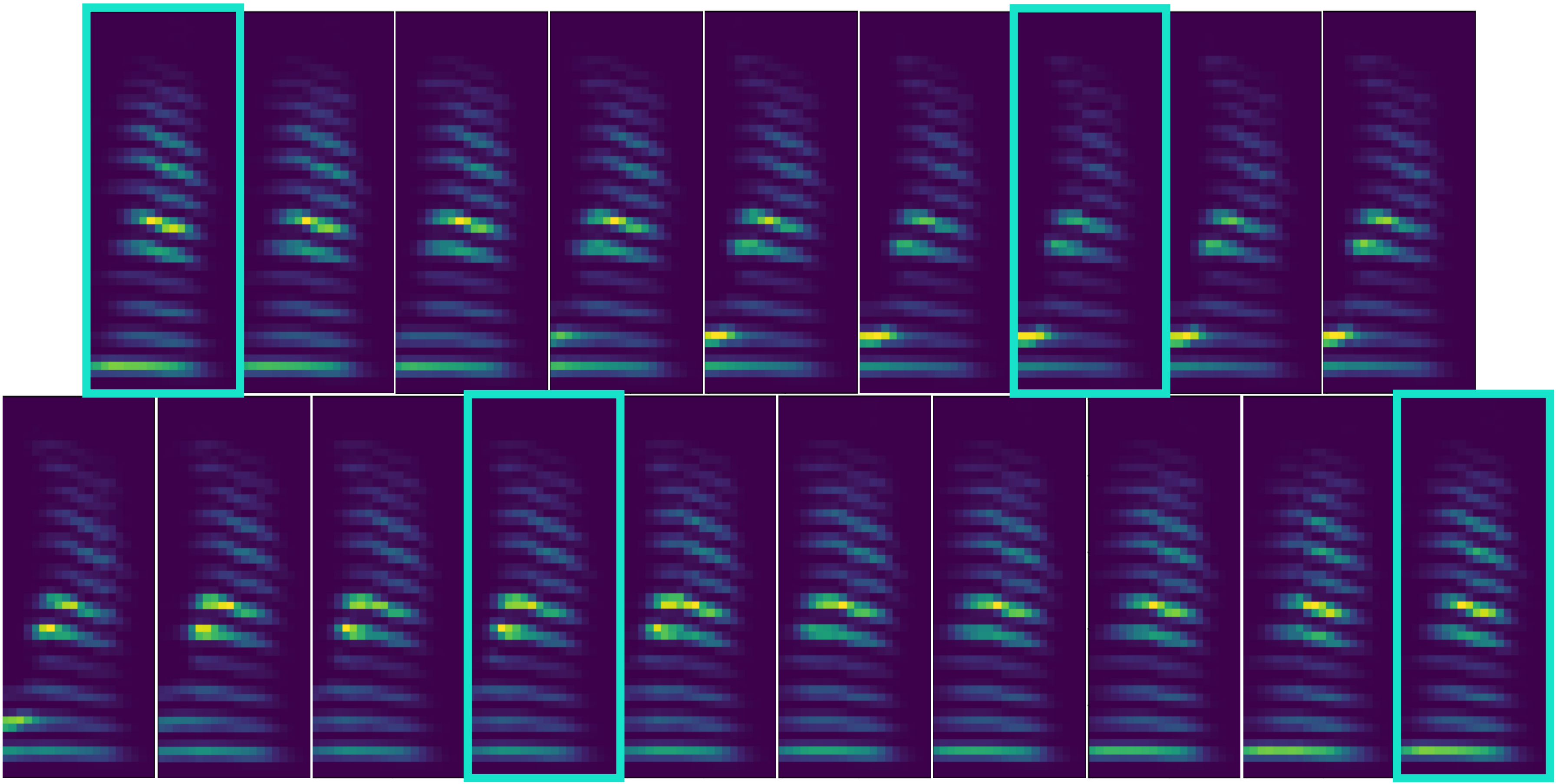}
  \caption{Alveolar consonant interpolation between ``ra'' (top left), ``na'', ``la'', and ``ra'' (bottom right), with these four diphones boxed.}
  \label{fig:alveolar_consonant_grad}
\end{figure}

\begin{figure}[t] 
  \includegraphics[width=38mm]{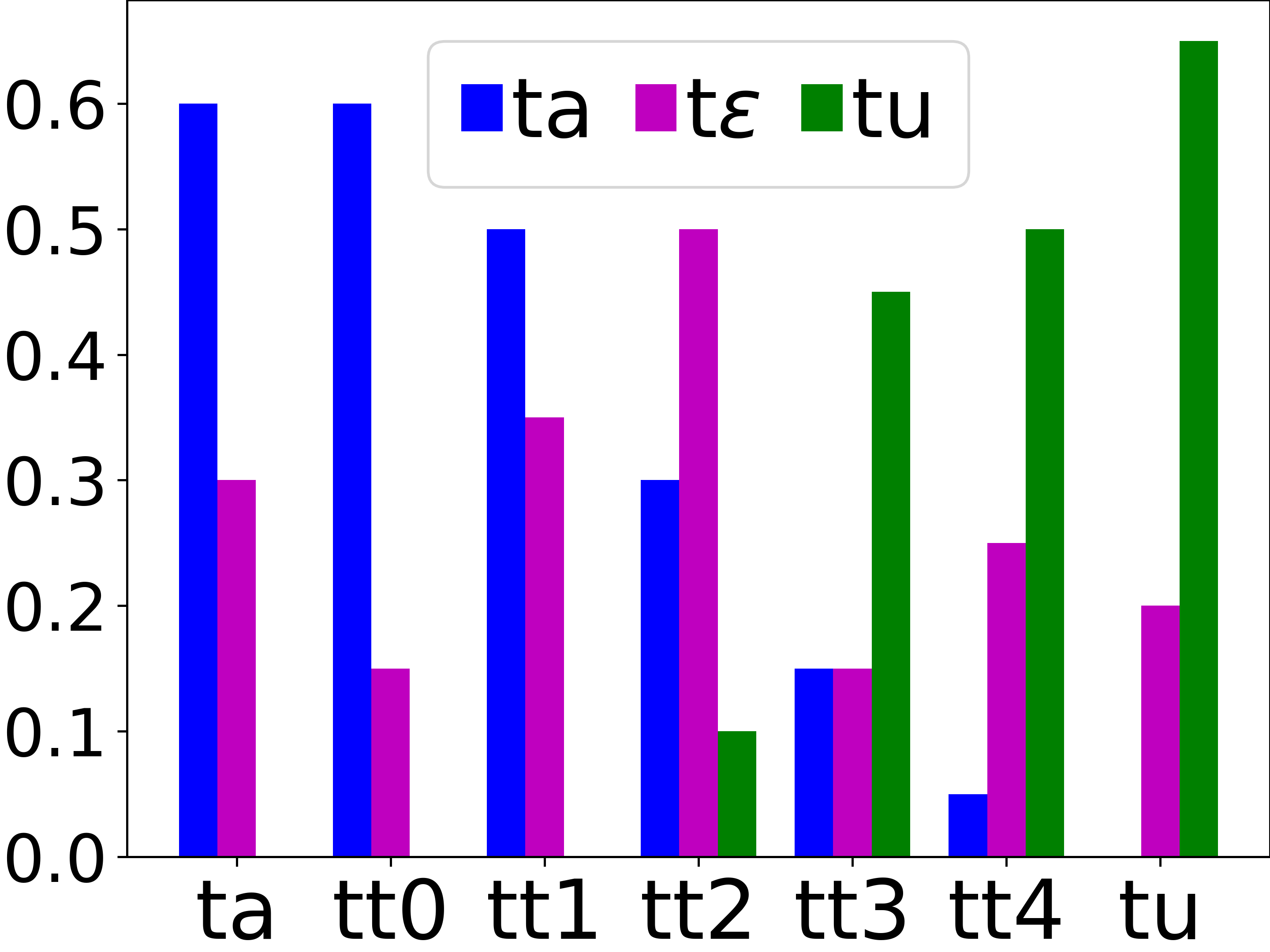}
  \includegraphics[width=38mm]{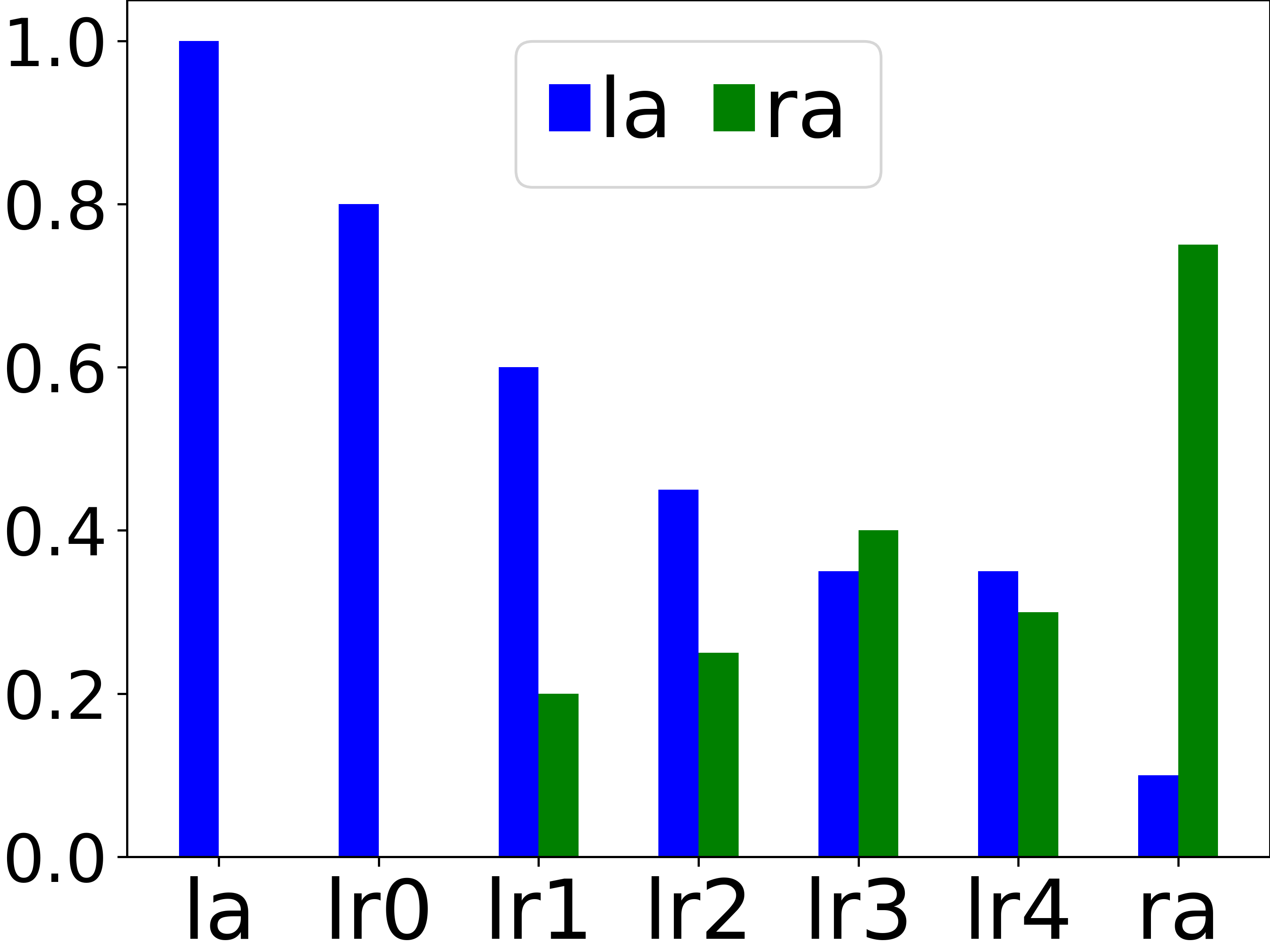}
  \caption{Human perception of ``ta''$\rightarrow$``tu'' (left) and ``la''$\rightarrow$``ra'' (right) interpolations. Details in Section \ref{sec:interpolation}.}
  \label{fig:interp_bar_plot}
\end{figure}

We also study the generalizability of our model with respect to consonants. To study how well our model generalizes across types of consonant sounds, we fix the place of articulation and interpolate between consonant types. 
Namely, we interpolate between the alveolar consonants "ra", "na", and "la", using the same generation and evaluation methodologies as our vowel interpolation experiment. Figure \ref{fig:alveolar_consonant_grad} depicts the mel-spectrograms of the synthesized interpolation samples. Our time-domain model smoothly transitions between these nasal, approximant, and lateral approximant consonants, as done in the human speech production process. MTurk listeners also perceived this behavior, based on Figure \ref{fig:interp_bar_plot} results similar to those in Section \ref{sec:vowel_interpolation}. Thus, while sounds between these diphones are not in our training set, our model is still able to generate unseen transitions that reflect those made by human articulators, suggesting an ability to generalize.

\subsection{Interpretability}

We note that these interpolation results also highlight the interpretability of articulatory features. Namely, we are able to simply take an element-wise weighted sum of two same-length sequences of articulatory features in order to create the utterance corresponding to articulator movements in between the two gestures. For example, to create the "t\textepsilon" sound, we would just need to synthesize the average of the articulatory feature sequences for "ti" and "ta". To our knowledge, this degree of interpretability is not supported by other speech representations like spectrums or deep-learning-based ones \cite{pmlr-v80-wang18h, sun2020taco}. 

\section{Synthesis Quality}
\label{sec:synthesis_quality}

\subsection{Subjective Fidelity Evaluation}

We first compare the fidelity our models by synthesizing the EMA-to-speech test set utterances and performing an AB naturalness preference test on MTurk with 40 total English-speaking listeners and 2 comparisons per datapoint, yielding 120 votes. Samples are available in the supplementary material linked in Section \ref{sec:intro}. Votes were evenly split between the models, with listeners agreeing on 28 out of 60 pairs, suggesting that listeners are unable to distinguish the two models. Thus, to study the differences between our models, we perform transcription and objective fidelity evaluations, as discussed below.

\subsection{Objective Fidelity Evaluation}
\label{sec:mcd}

In order to perform an objective evaluation of synthesis quality, we compute the mel-cepstral distortion (MCD) \cite{black2019cmu_wilderness} between the ground truth and the generations from both models for each utterance in our test sets. As described in Table \ref{mcd}, our time-domain articulatory synthesis approach performs better than the spectrum-intermediate baseline on the synthetic articulatory dataset and slightly worse on the EMA-to-speech task. We attribute the performance drop of our model on the EMA task to information loss within in the input data. Namely, the model appears to confuse phonemes due to the lack of manner information in the EMA inputs, as heard in the accompanying samples and analyzed in Section \ref{sec:phoneme_confusion}.

\begin{table}[t]
\centering
\begin{tabular}{lcc}
\hline
\textbf{Model} & \multicolumn{2}{c}{\textbf{MCD} \textcolor{ForestGreen}{$\downarrow$}} \\
{} & \textbf{Birk.-Pseudo} & \textbf{EMA-MGNU0}\\
\hline
HiFi-CAR & {$\mathbf{3.26 \pm 0.18}$} & ${4.81 \pm 0.73}$ \\
Spec.-Int. & {$5.15 \pm 0.48$} & {$\mathbf{4.75 \pm 0.81}$} \\
\hline
\end{tabular}
\caption{\label{mcd}
MCD for each model on Birkholz and EMA data. See Section \ref{sec:mcd} for details.
}
\end{table}

\subsection{Transcription}
\label{sec:transcription}

To evaluate intelligibility, we conduct open-vocabulary transcription experiments using the EMA-to-speech task, for which the test set contains non-synthetic utterances. First, we perform an objective automatic speech recognition (ASR) evaluation using DeepSpeech \cite{Hannun2014DeepSpeech} and Wav2Vec2 \cite{baevski2020wav2vec2, speechbrain}. We use ASR to transcribe the synthesis outputs of our models on the entire MNGU0 evaluation set described in Section \ref{sec:ema_mngu0_dataset} and calculate the average word error rates (WERs). We also evaluate the intelligibility of our model though human evaluations, following the MTurk methodology in Taguchi and Kaburagi's EMA-to-speech work \cite{taguchi2018articulatory} in order to compare with their WER 30.1\% result. Table \ref{tab:wer} summarizes our WERs. Based on 120 transcriptions per model by a total of 70 English-speaking listeners, our models achieve an average WER between 18\% and 19\%, i.e., improving WER by over 11\%. Our models have comparable WERs for both the objective and human evaluations, suggesting that they are both suitable for generating intelligible speech.

\begin{table}[t]
\centering
\begin{tabular}{lccc}
\hline
\textbf{Model} & \multicolumn{3}{c}{\textbf{WER} \textcolor{ForestGreen}{$\downarrow$}} \\
{} & \textbf{Human} & \textbf{DeepSpeech} & \textbf{Wav2Vec2}\\
\hline
HiFi-CAR & {$\mathbf{18.5}$} & ${34.7}$ & {$\mathbf{21.7}$} \\
Spec.-Int. & {$ 18.9$} & {$\mathbf{31.0}$} & ${23.5}$ \\
\hline
\end{tabular}
\caption{\label{tab:wer}
WER for each model. See Section \ref{sec:transcription} for details.
}
\end{table}

\section{Phoneme Confusion}
\label{sec:phoneme_confusion}

To study the phonological errors made by our model, we analyze the phonemes that our time-domain EMA-to-speech model confused during synthesis. For all ASR-transcribed test utterances and their ground truth texts, we convert the graphemes to phoneme sequences using Phonemizer \cite{Bernard2021phonemizer}. We identify phoneme confusion pairs using sclite,\footnote{\href{https://github.com/usnistgov/SCTK}{https://github.com/usnistgov/SCTK}} which aligns each predicted sequence with the respective ground truth and then records the substitution errors. Figure \ref{fig:asr_confusion} plots pairs that are confused more than once on an International Phonetic Alphabet (IPA) chart \cite{gaddy-klein-2021-improved}, where pairs with higher frequencies of substitution errors have darker lines. Most of the word substitution errors are due to plosive or vowel confusions. One potential reason for the plosive substitutions is that plosives generally have a shorter duration than other consonant types like fricatives \cite{alwan2011plosivefricative} and thus may be more readily confusable. Multiple voiced-unvoiced pairs are also confused, which may be because estimated F0 is the only voicing information in the input, as described in Section \ref{sec:ema_mngu0_dataset}. These results suggest that future work adding additional articulatory features like velar and glottal information can yield even higher fidelity articulatory synthesizers, potentially comparable to current text-to-speech (TTS) systems.

\pgfdeclarelayer{background}
\pgfsetlayers{background,main}
\let\ipa\textipa
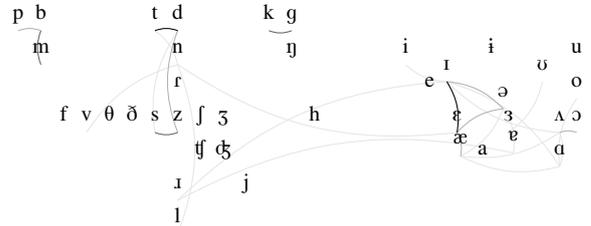
\begin{figure}[t]
  \centering
  \begin{tikzpicture}[x=.5cm,y=-.75cm, scale=.6, every node/.style={scale=.9,font=\vphantom{pb}}]
  \node at (-2, 0) (sil) {};
   
  \begin{scope}[shift={(-7,0)}]
    \node at (1, 0) (p) {p};
    \node at (2, 0) (b) {b};
    \node at (7, 0) (t) {t};
    \node at (8, 0) (d) {d};
    \node at (12, 0) (k) {k};
    \node at (13, 0) (g) {\ipa{g}};
    \node at (2, 1) (m) {m};
    \node at (8, 1) (n) {n};
    \node at (13, 1) (ng) {\ipa{N}};
    \node at (8, 2) (R) {\ipa{R}};
    \node at (3, 3) (f) {f};
    \node at (4, 3) (v) {v};
    \node at (5, 3) (th) {\ipa{T}};
    \node at (6, 3) (dh) {\ipa{D}};
    \node at (7, 3) (s) {s};
    \node at (8, 3) (z) {z};
    \node at (9, 3) (sh) {\ipa{S}};
    \node at (10, 3) (zh) {\ipa{Z}};
    \node at (14, 3) (hh) {h};
    \node at (9, 4) (ch) {\textteshlig};
    \node at (10,4) (jh) {\textdyoghlig};
    \node at (8,5) (*r) {\textturnr};
    \node at (11,5) (j) {j};
    \node at (8,6) (l) {l};
  \end{scope}
   
  \begin{scope}[shift={(20,1)}, x=.75cm]
  \node at (-1, 0) (uw) {u};
  \node at (-1, 1) (o) {o};
  \node at (-2, .5) (uh) {\ipa{U}};
  \node at (-1, 2) (ao) {\ipa{O}};
  \node at (-1.5, 2) (ah) {\ipa{2}};
  \node at (-1.5, 3) (aa) {\ipa{A}};
  \node at (-3.15, 1.3) (ax) {\ipa{@}};
  \node at (-6, 0) (iy) {i};
  \node at (-4.8, .5) (ih) {\ipa{I}};
  \node at (-5.3, 1) (e) {e};
  \node at (-3.5, 0) (1) {\ipa{1}};
  \node at (-4.5, 2) (eh) {\ipa{E}};
  \node at (-3, 2) (3) {\ipa{3}};
  \node at (-2.85, 2.6) (5) {\ipa{5}};
  \node at (-4.4, 2.7) (ae) {\ipa{\ae}};
  \node at (-3.75, 3) (a) {a};
  \end{scope}
   
\begin{pgfonlayer}{background}


\draw[color=black!10.000000] (ih.south) to[bend right=20] (ah.south); 
\draw[color=black!10.000000] (ih.south) to[bend right=20] (*r.south); 
\draw[color=black!10.000000] (3.south) to[bend right=20] (uh.south); 
\draw[color=black!10.000000] (aa.south) to[bend right=20] (ah.south); 
\draw[color=black!10.000000] (aa.south) to[bend right=20] (ih.south); 
\draw[color=black!10.000000] (5.south) to[bend right=20] (*r.south); 
\draw[color=black!10.000000] (5.south) to[bend right=20] (ax.south); 
\draw[color=black!10.000000] (ae.south) to[bend right=20] (ah.south); 
\draw[color=black!10.000000] (ae.south) to[bend right=20] (ax.south); 
\draw[color=black!10.000000] (ae.south) to[bend right=20] (aa.south); 
\draw[color=black!10.000000] (o.south) to[bend right=20] (aa.south); 
\draw[color=black!10.000000] (n.south) to[bend right=20] (eh.south); 
\draw[color=black!10.000000] (n.south) to[bend right=20] (v.south); 
\draw[color=black!10.000000] (n.south) to[bend right=20] (t.south); 
\draw[color=black!10.000000] (l.south) to[bend right=20] (n.south); 
\draw[color=black!10.000000] (iy.south) to[bend right=20] (ih.south); 
\draw[color=black!10.000000] (d.south) to[bend right=20] (s.south); 
\draw[color=black!30.000000] (ax.south) to[bend right=20] (ih.south); 
\draw[color=black!30.000000] (ax.south) to[bend right=20] (eh.south); 
\draw[color=black!30.000000] (ao.south) to[bend right=20] (ah.south); 
\draw[color=black!30.000000] (ae.south) to[bend right=20] (eh.south); 
\draw[color=black!30.000000] (d.south) to[bend right=20] (z.south); 
\draw[color=black!30.000000] (b.south) to[bend right=20] (p.south); 
\draw[color=black!60.000000] (s.south) to[bend right=20] (z.south); 
\draw[color=black!60.000000] (k.south) to[bend right=20] (g.south); 
\draw[color=black!60.000000] (b.south) to[bend right=20] (m.south); 
\draw[color=black!120.000000] (eh.south) to[bend right=20] (ih.south); 
\draw[color=black!150.000000] (d.south) to[bend right=20] (t.south); 

\end{pgfonlayer}
   
  \end{tikzpicture}
  \caption{Phoneme confusability based on ASR transcriptions. Phoneme pairs that are confused more frequently have darker lines. Details in Section \ref{sec:phoneme_confusion}.}
  \label{fig:asr_confusion}
\end{figure}

\section{Conclusion and Future Directions}

In this work, we study ways to build deep articulatory synthesizers that are efficient and high-fidelity. Based on computational efficiency evaluations, we observe that our proposed time-domain methodology is suitable for achieving time and space complexities that are noticeably lower than the baseline spectrum-intermediate approach. Our interpolation study also highlights the generalizability and interpretability of our approach. Through MCD, ASR, and human transcription experiments, we demonstrate the intelligibility of our model, improving the transcription word error rate for the EMA-to-speech task by over 11\%. Moving forward, we plan to test our methodology on other modalities like electromyography \cite{gaddy-klein-2021-improved} and real-time magnetic resonance imaging \cite{lim2021mri}.

\section{Acknowledgements}

This research is supported by the National Science Foundation (Award 2106928).

\bibliographystyle{IEEEtran}

\bibliography{mybib}

\begin{thebibliography}{10}
\providecommand{\url}[1]{#1}
\csname url@samestyle\endcsname
\providecommand{\newblock}{\relax}
\providecommand{\bibinfo}[2]{#2}
\providecommand{\BIBentrySTDinterwordspacing}{\spaceskip=0pt\relax}
\providecommand{\BIBentryALTinterwordstretchfactor}{4}
\providecommand{\BIBentryALTinterwordspacing}{\spaceskip=\fontdimen2\font plus
\BIBentryALTinterwordstretchfactor\fontdimen3\font minus
  \fontdimen4\font\relax}
\providecommand{\BIBforeignlanguage}[2]{{%
\expandafter\ifx\csname l@#1\endcsname\relax
\typeout{** WARNING: IEEEtran.bst: No hyphenation pattern has been}%
\typeout{** loaded for the language `#1'. Using the pattern for}%
\typeout{** the default language instead.}%
\else
\language=\csname l@#1\endcsname
\fi
#2}}
\providecommand{\BIBdecl}{\relax}
\BIBdecl

\bibitem{Wang2017Tacotron}
Y.~Wang \emph{et~al.}, ``Tacotron: Towards end-to-end speech synthesis,'' in
  \emph{Interspeech}, 2017.

\bibitem{hayashi2021espnet2}
T.~Hayashi \emph{et~al.}, ``Espnet2-tts: Extending the edge of tts research,''
  \emph{arXiv preprint arXiv:2110.07840}, 2021.

\bibitem{prenger2019waveglow}
R.~Prenger, R.~Valle, and B.~Catanzaro, ``Waveglow: A flow-based generative
  network for speech synthesis,'' in \emph{ICASSP}, 2019.

\bibitem{tjandra2019speechtranslation}
A.~Tjandra, S.~Sakti, and S.~Nakamura, ``Speech-to-speech translation between
  untranscribed unknown languages,'' in \emph{ASRU}, 2019, pp. 593--600.

\bibitem{jia2019speechtranslation}
Y.~Jia \emph{et~al.}, ``Direct speech-to-speech translation with a
  sequence-to-sequence model,'' in \emph{Interspeech}, 2019.

\bibitem{inaguma2020espnetspeechtranslation}
H.~Inaguma \emph{et~al.}, ``{ESP}net-{ST}: All-in-one speech translation
  toolkit,'' in \emph{ACL}, 2020.

\bibitem{polyak21facebookresynthesis}
A.~Polyak \emph{et~al.}, ``{Speech Resynthesis from Discrete Disentangled
  Self-Supervised Representations},'' in \emph{Interspeech}, 2021.

\bibitem{wu2021privacy}
P.~Wu \emph{et~al.}, ``Understanding the tradeoffs in client-side privacy for
  downstream speech tasks,'' in \emph{APSIPA ASC}, 2021.

\bibitem{sisman2020overview}
B.~Sisman \emph{et~al.}, ``An overview of voice conversion and its challenges:
  From statistical modeling to deep learning,'' \emph{TASLP}, 2020.

\bibitem{anumanchipalli2019speech}
G.~K. Anumanchipalli, J.~Chartier, and E.~F. Chang, ``Speech synthesis from
  neural decoding of spoken sentences,'' \emph{Nature}, vol. 568, no. 7753, pp.
  493--498, 2019.

\bibitem{yu2019durianmultimodalsynthesis}
C.~Yu \emph{et~al.}, ``Durian: Duration informed attention network for
  multimodal synthesis,'' \emph{arXiv preprint arXiv:1909.01700}, 2019.

\bibitem{gaddy-klein-2021-improved}
D.~Gaddy and D.~Klein, ``An improved model for voicing silent speech,'' in
  \emph{ACL-IJCNLP}, 2021.

\bibitem{karmel2019speechsynthesisforblind}
A.~Karmel \emph{et~al.}, ``Iot based assistive device for deaf, dumb and blind
  people,'' \emph{ICRTAC}, 2019.

\bibitem{angrick2019ecogbraintospeech}
M.~Angrick \emph{et~al.}, ``Speech synthesis from ecog using densely connected
  3d convolutional neural networks,'' \emph{Journal of neural engineering},
  vol.~16, no.~3, p. 036019, 2019.

\bibitem{jia2021translatotron}
Y.~Jia \emph{et~al.}, ``Translatotron 2: Robust direct speech-to-speech
  translation,'' \emph{arXiv preprint arXiv:2107.08661}, 2021.

\bibitem{Nekvinda2020multilingualtts}
T.~Nekvinda and O.~Dušek, ``{One Model, Many Languages: Meta-Learning for
  Multilingual Text-to-Speech},'' in \emph{Interspeech}, 2020.

\bibitem{zhang2019multilingualtts}
Y.~Zhang \emph{et~al.}, ``Learning to speak fluently in a foreign language:
  Multilingual speech synthesis and cross-language voice cloning,'' \emph{arXiv
  preprint arXiv:1907.04448}, 2019.

\bibitem{fant1991articulatorysynthesis}
G.~Fant, ``What can basic research contribute to speech synthesis?''
  \emph{Journal of Phonetics}, vol.~19, no.~1, pp. 75--90, 1991.

\bibitem{rubin1981articulatorysynthesis}
P.~Rubin, T.~Baer, and P.~Mermelstein, ``An articulatory synthesizer for
  perceptual research,'' \emph{The Journal of the Acoustical Society of
  America}, vol.~70, no.~2, pp. 321--328, 1981.

\bibitem{scully1990articulatorysynthesis}
C.~Scully, ``Articulatory synthesis,'' in \emph{Speech production and speech
  modelling}.\hskip 1em plus 0.5em minus 0.4em\relax Springer, 1990, pp.
  151--186.

\bibitem{jiachen2022gesture}
J.~Lian, A.~W. Black, L.~Goldstein, and G.~K. Anumanchipalli, ``Deep neural
  convolutive matrix factorization for articulatory representation
  decomposition,'' \emph{Interspeech}, 2022.

\bibitem{fant1995lfvocaltractmodal}
G.~Fant, ``The lf-model revisited. transformations and frequency domain
  analysis,'' \emph{Speech Trans. Lab. Q. Rep., Royal Inst. of Tech.
  Stockholm}, vol.~2, no.~3, p.~40, 1995.

\bibitem{iskarous2003haskinscasy}
K.~Iskarous \emph{et~al.}, ``Casy: The haskins configurable articulatory
  synthesizer,'' in \emph{International Congress of Phonetic Sciences}, 2003.

\bibitem{birkholz2013vtl}
P.~Birkholz, ``Modeling consonant-vowel coarticulation for articulatory speech
  synthesis,'' \emph{PloS one}, vol.~8, no.~4, p. e60603, 2013.

\bibitem{aryal2016datadrivendnn}
S.~Aryal and R.~Gutierrez-Osuna, ``Data driven articulatory synthesis with deep
  neural networks,'' \emph{Computer Speech \& Language}, vol.~36, pp. 260--273,
  2016.

\bibitem{bocquelet2014dnn}
F.~Bocquelet \emph{et~al.}, ``Robust articulatory speech synthesis using deep
  neural networks for bci applications,'' in \emph{Interspeech}, 2014.

\bibitem{chen2021ema2s}
Y.-W. Chen \emph{et~al.}, ``Ema2s: An end-to-end multimodal
  articulatory-to-speech system,'' in \emph{ISCAS}, 2021.

\bibitem{kong2020hifigan}
J.~Kong, J.~Kim, and J.~Bae, ``Hifi-gan: Generative adversarial networks for
  efficient and high fidelity speech synthesis,'' in \emph{NeurIPS}, 2020.

\bibitem{morrison2022cargan}
M.~Morrison \emph{et~al.}, ``Chunked autoregressive gan for conditional
  waveform synthesis,'' in \emph{Submitted to ICLR 2022}, April 2022.

\bibitem{badlani2021onettsalignment}
R.~Badlani \emph{et~al.}, ``One tts alignment to rule them all,'' \emph{arXiv
  preprint arXiv:2108.10447}, 2021.

\bibitem{kim2021vits}
J.~Kim, J.~Kong, and J.~Son, ``Conditional variational autoencoder with
  adversarial learning for end-to-end text-to-speech,'' in \emph{ICML}, 2021.

\bibitem{Elias2021ParallelTacotron2}
I.~Elias \emph{et~al.}, ``Parallel tacotron 2: A non-autoregressive neural tts
  model with differentiable duration modeling,'' \emph{ArXiv}, vol.
  abs/2103.14574, 2021.

\bibitem{nam2012procedure}
H.~Nam \emph{et~al.}, ``A procedure for estimating gestural scores from speech
  acoustics,'' \emph{The Journal of the Acoustical Society of America}, vol.
  132, no.~6, pp. 3980--3989, 2012.

\bibitem{richmond2011mngu0}
K.~Richmond, P.~Hoole, and S.~King, ``Announcing the electromagnetic
  articulography (day 1) subset of the mngu0 articulatory corpus,'' in
  \emph{Interspeech}, 08 2011, pp. 1505--1508.

\bibitem{taguchi2018articulatory}
F.~Taguchi and T.~Kaburagi, ``Articulatory-to-speech conversion using
  bi-directional long short-term memory.'' in \emph{Interspeech}, 2018, pp.
  2499--2503.

\bibitem{stone2020ema}
S.~Stone, P.~Schmidt, and P.~Birkholz, ``Prediction of voicing and the f0
  contour from electromagnetic articulography data for articulation-to-speech
  synthesis,'' in \emph{ICASSP}, 2020.

\bibitem{liu2018articulatory}
Z.-C. Liu, Z.-H. Ling, and L.-R. Dai, ``Articulatory-to-acoustic conversion
  using blstm-rnns with augmented input representation,'' \emph{Speech
  Communication}, vol.~99, pp. 161--172, 2018.

\bibitem{Csapo2020UltrasoundbasedAM}
T.~G. Csap'o \emph{et~al.}, ``Ultrasound-based articulatory-to-acoustic mapping
  with waveglow speech synthesis,'' in \emph{Interspeech}, 2020.

\bibitem{georges2020towards}
M.-A. Georges \emph{et~al.}, ``Towards an articulatory-driven neural vocoder
  for speech synthesis,'' in \emph{International Seminar on Speech Production},
  2020.

\bibitem{Vaswani2017attentiontransformer}
A.~Vaswani \emph{et~al.}, ``Attention is all you need,'' in \emph{NeurIPS},
  2017.

\bibitem{kim2018crepe}
J.~W. Kim \emph{et~al.}, ``Crepe: A convolutional representation for pitch
  estimation,'' in \emph{ICASSP}, 2018.

\bibitem{hagedorn11_interspeech}
C.~Hagedorn, M.~Proctor, and L.~Goldstein, ``{Automatic analysis of singleton
  and geminate consonant articulation using real-time magnetic resonance
  imaging},'' in \emph{Interspeech}, 2011.

\bibitem{lenzo2003kal}
K.~Lenzo and A.~Black, ``Diphone collection and synthesis,'' \emph{ICSLP},
  2000.

\bibitem{alfonso1982dynamics}
P.~J. Alfonso and T.~Baer, ``Dynamics of vowel articulation,'' \emph{Language
  and Speech}, vol.~25, no.~2, pp. 151--173, 1982.

\bibitem{pmlr-v80-wang18h}
Y.~Wang \emph{et~al.}, ``Style tokens: Unsupervised style modeling, control and
  transfer in end-to-end speech synthesis,'' in \emph{ICML}, 2018.

\bibitem{sun2020taco}
G.~Sun \emph{et~al.}, ``Fully-hierarchical fine-grained prosody modeling for
  interpretable speech synthesis,'' in \emph{ICASSP}, 2020.

\bibitem{black2019cmu_wilderness}
A.~W. Black, ``{CMU} wilderness multilingual speech dataset,'' in
  \emph{ICASSP}, 2019.

\bibitem{Hannun2014DeepSpeech}
A.~Y. Hannun \emph{et~al.}, ``Deep speech: Scaling up end-to-end speech
  recognition,'' \emph{ArXiv}, vol. abs/1412.5567, 2014.

\bibitem{baevski2020wav2vec2}
A.~Baevski \emph{et~al.}, ``Wav2vec 2.0: A framework for self-supervised
  learning of speech representations,'' in \emph{NeurIPS}, 2020.

\bibitem{speechbrain}
M.~Ravanelli \emph{et~al.}, ``{SpeechBrain}: A general-purpose speech
  toolkit,'' 2021, arXiv:2106.04624.

\bibitem{Bernard2021phonemizer}
M.~Bernard and H.~Titeux, ``Phonemizer: Text to phones transcription for
  multiple languages in python,'' \emph{Journal of Open Source Software},
  vol.~6, no.~68, p. 3958, 2021.

\bibitem{alwan2011plosivefricative}
A.~Alwan, J.~Jiang, and W.~Chen, ``Perception of place of articulation for
  plosives and fricatives in noise,'' \emph{Speech communication}, vol.~53,
  no.~2, pp. 195--209, 2011.

\bibitem{lim2021mri}
Y.~Lim \emph{et~al.}, ``A multispeaker dataset of raw and reconstructed speech
  production real-time mri video and 3d volumetric images,'' \emph{Scientific
  Data}, vol.~8, 07 2021.

\end{thebibliography}


\end{document}